\documentclass[conference]{IEEEtran}
\usepackage{amsmath,amssymb,amsfonts}
\usepackage{algorithm,algorithmic}
\usepackage{array}
\usepackage{cite}
\usepackage{textcomp}
\usepackage{stfloats}
\usepackage{setspace}
\usepackage{url}
\usepackage{xcolor}
\usepackage{multirow}
\usepackage{verbatim}
\usepackage{graphicx}
\usepackage{diagbox}
\usepackage{subfig}
\usepackage{hyperref}
\usepackage{bm}
\usepackage{xcolor}
\hypersetup{
	colorlinks=true,
	linkcolor=blue,
	filecolor=blue,
	urlcolor=blue,
	citecolor=cyan,
}
\usepackage{balance}
\begin{document}
	\columnsep 0.205 in
	
	\title{Coherence-Minimized Sensing Matrix Design for MRI Reconstruction via Dual-Space Projection Optimization}
	\author{
		\IEEEauthorblockN{Siyuan Feng}
		\IEEEauthorblockA{
			School of Computer Science and Engineering, 
			Northeastern University, Shenyang, China \\
			20236189@stu.neu.edu.cn
		}
	}
	
	\maketitle
	\begin{abstract}
	Compressed sensing magnetic resonance imaging (CS-MRI) heavily relies on the low mutual coherence between the measurement matrix and the sparsity basis. 
	However, under highly accelerated Cartesian undersampling, 
	the severe structural coherence between Fourier measurements and spatial bases, discrete cosine transform (DCT) for example, fundamentally violates this requirement, causing classical sparse recovery algorithms to stagnate. 
	To mitigate this fundamental bottleneck, we propose a synergistic dual-space projection framework, denoted as $\mathbf{PAQ}$. 
	Instead of merely designing heuristic sampling masks, our method directly reshapes the equivalent dictionary. 
	Specifically, we introduce a diagonal-dominant random rotator $\mathbf{Q}$ in the feature space to probabilistically disrupt structural alignment, 
	and an active orthogonalization projector $\mathbf{P}$ in the measurement space to deterministically whiten the residual correlations. 
	We theoretically demonstrate that this dual-space mechanism bounds the mutual coherence with an exponentially decaying tail via sub-exponential distribution properties. 
	Experimental validations on clinical MRI datasets under 20\% Cartesian sampling demonstrate that plugging the $\mathbf{PAQ}$ 
	preconditioners into the standard ISTA solver significantly suppresses aliasing artifacts and yields consistent peak signal-to-noise ratio (PSNR) improvements.
	\end{abstract}

	\begin{IEEEkeywords}
		Sparse Signal Recovery, Compressed Sensing, MRI Reconstruction, Mutual Coherence
	\end{IEEEkeywords}

	\section{Introduction}
	    Magnetic resonance imaging (MRI) is a ubiquitous non-invasive diagnostic modality, yet its inherently slow acquisition speed limits clinical throughput and induces motion artifacts.

    Compressed sensing provides a mathematical foundation to accelerate MRI by recovering high-fidelity images from significantly undersampled k-space measurements \cite{lustig2008}.

    This is theoretically guaranteed provided that the underlying image is sparse in a certain transform domain, wavelet \cite{mallat1989} and DCT for example, and the measurement matrix satisfy the restricted isometry property \cite{candes2006rip} or possesses low mutual coherence.

	In clinical practice, Cartesian sampling trajectories are heavily favored due to their robustness to hardware imperfections. However, highly accelerated Cartesian undersampling 
	(e.g., retaining only 20\% of the data) introduces severe structural coherence between the Fourier measurement operator and the sparsity basis, 
	discrete this ill-conditioning fundamentally breaks the mathematical guarantees of standard CS, causing classical optimization algorithms, 
	such as the iterative soft-thresholding algorithm (ISTA) \cite{sui2024} 
	and proximal methods \cite{Dimas2024}, to suffer from severe performance degradation and structured aliasing artifacts.

	To alleviate this coherence bottleneck, substantial efforts have been devoted to optimizing the sensing dictionary \cite{aharon2006, Heesterbeek2024}. Specifically, optimizing the measurement matrix has emerged as a promising direction. 
	For instance, Bruckstein \cite{bruckstein2008} theoretically demonstrated that left-multiplying the linear sensing system by a well-designed conditioning matrix can effectively reduce mutual coherence and improve the bounds for exact sparse recovery. 
	Similarly, Abolghasemi \cite{abolghasemi2010} proposed minimizing the Frobenius norm of the Gram matrix's off-diagonal elements via gradient descent to yield a nearly orthogonal measurement system. 
	While mathematically elegant, directly applying these state-of-the-art conditioning methods to CS-MRI requires explicit computation of the $N \times N$ equivalent dictionary $\mathbf{D} = \mathbf{A}\boldsymbol{\Psi}$, which is computationally prohibitive (typically $\mathcal{O}(N^2)$) for high-resolution clinical images. 

	To bypass these computational bottlenecks, we propose a fundamentally different approach: 
	the \textbf{Dual-Space PAQ Optimization}. Instead of merely optimizing heuristic sampling masks, 
	our framework directly reshapes the equivalent dictionary by synergizing two specialized operators. 
	Specifically, we introduce an active orthogonalization projector $\mathbf{P}$ in the measurement space to deterministically whiten residual correlations, 
	coupled with a diagonal-dominant random rotator $\mathbf{Q}$ in the feature space \cite{dogrot2026} to probabilistically disrupt structural alignment. 
	This dual-space mechanism effectively minimizes mutual coherence, 
	providing a robust preconditioning framework for various CS-MRI reconstruction tasks.

	The remainder of this paper is organized as follows: Section II reviews the background. 
	Section III details the construction of the dual-space $\mathbf{PAQ}$ framework. 
	Section IV provides theoretical analysis. Section V presents experimental validations, followed by the conclusion in Section VI.

	\section{Background and Problem Formulation}
	\label{sec:background}

	To establish the theoretical foundation for the proposed dual-space preconditioning, 
	this section briefly reviews the standard Compressed Sensing MRI mathematical model and analyzes the fundamental bottleneck caused by mutual coherence.

	\subsection{CS-MRI Model and Coherence}
	Standard CS-MRI aims to reconstruct a sparse image $\mathbf{x} \in \mathbb{C}^N$ from undersampled $k$-space measurements 
	$\mathbf{y} = \mathbf{A}\mathbf{x} + \mathbf{n}$, where $\mathbf{A} = \mathbf{M}\mathbf{F}$ is the measurement matrix 
	consisting of a Fourier transform $\mathbf{F}$ and an undersampling mask $\mathbf{M}$. 
	The reconstruction is typically formulated as a sparse regularization problem: $\min_{\mathbf{x}} \|\boldsymbol{\Psi}\mathbf{x}\|_1 \text{ s.t. } \|\mathbf{y}-\mathbf{A}\mathbf{x}\|_2 < \sigma$, 
	where $\boldsymbol{\Psi}$ denotes the sparsity basis. According to CS theory, 
	the performance of this recovery is fundamentally limited by the mutual coherence between $\mathbf{A}$ and $\boldsymbol{\Psi}$.

	\subsection{The Mutual Coherence Bottleneck}
	The theoretical guarantee of successfully solving the sparse recovery problem 
	heavily relies on the properties of the equivalent dictionary $\mathbf{D} = \mathbf{A}\boldsymbol{\Psi} \in \mathbb{C}^{M \times N}$. A fundamental metric evaluating the suitability of $\mathbf{D}$ is the mutual coherence, defined as the maximum absolute inner product between any two normalized columns $\mathbf{d}_i$ and $\mathbf{d}_j$:
	\begin{equation}
		\mu(\mathbf{D}) = \max_{1 \le i \neq j \le N} \frac{|\mathbf{d}_i^H \mathbf{d}_j|}{\|\mathbf{d}_i\|_2 \|\mathbf{d}_j\|_2}
		\label{eq:coherence}
	\end{equation}
	The mutual coherence $\mu(\mathbf{D}) \in [0, 1]$ dictates the upper bound of the sparsity level that can be reliably recovered. A smaller $\mu$ indicates that the columns of the dictionary are nearly orthogonal, which prevents the sparse solver from confusing different structural components.

	However, a severe bottleneck arises under aggressive Cartesian undersampling. 
	Because Cartesian masks predominantly sample the low-frequency center of the k-space to capture global image contrast, 
	the rows of $\mathbf{A}$ become highly structured. 
	When coupled with spatial bases like DCT, the resulting dictionary $\mathbf{D}$ exhibits extreme structural alignment. 
	This drives the mutual coherence $\mu(\mathbf{D})$ dangerously close to $1$, 
	severely violating the restricted isometry property (RIP). 
	Consequently, the standard ISTA solver fails to isolate the true sparse coefficients, 
	leading to the structured aliasing artifacts observed in classical CS-MRI reconstructions. 

	This inherent mathematical limitation motivates the necessity of explicitly restructuring the sensing environment, 
	rather than merely adjusting the numerical solver, leading to our proposed $\mathbf{PAQ}$ framework.

	\section{Proposed Method: Dual-Space PAQ Optimization}

	\subsection{Feature-Space Decorrelation via Random Rotator}
	\label{sec:q_matrix}
	
	{
	
	To disrupt the structural coherence between the measurement matrix and the sparsity basis, we introduce a diagonal-dominant random rotator $\mathbf{Q} \in \mathbb{R}^{N \times N}$ in the feature space. Following the DogRot framework \cite{dogrot2026}, $\mathbf{Q}$ is constructed as:
	\begin{equation}
		\mathbf{Q} = \mathbf{I}_N + \mathbf{\Delta} - \mathbf{\Delta}^T
		\label{eq:q_construct}
	\end{equation}
	where $\mathbf{I}_N$ is the identity matrix, and $\mathbf{\Delta}$ is a strictly lower-triangular matrix with entries independently drawn from $\mathcal{N}(0, \epsilon^2)$. 

	This construction ensures two key properties:
	\begin{enumerate}
		\item \textbf{Sparsity Preservation:} Since $\epsilon$ is small, $\mathbf{Q}$ acts as a near-identity operator, which minimally affects the signal's sparsity support.
		\item \textbf{Random Decorrelation:} The skew-symmetric component $\mathbf{\Delta} - \mathbf{\Delta}^T$ introduces random perturbations that effectively misalign the dictionary atoms from the coherent directions of the sampling operator, thereby reducing mutual coherence.
	\end{enumerate}
	}

	\subsection{Measurement-Space Active Orthogonalization}

	While $\mathbf{Q}$ injects controlled randomness to disrupt structural coherence, it does not explicitly minimize it. 
	We achieve this through active orthogonalization in the measurement space.

	\subsubsection{The Unitary Simplification and Dual Matrix}
	To simplify the computation of the dual Gram matrix $\mathbf{B} = \mathbf{D}\mathbf{D}^H$, we leverage the fact that the sparsity basis $\boldsymbol{\Psi}$ is a unitary operator. Consequently, $\boldsymbol{\Psi}\boldsymbol{\Psi}^H = \mathbf{I}_N$, allowing $\mathbf{B}$ to be reduced to a measurement-space operator:
	\begin{equation}
		\mathbf{B} = \mathbf{A}\boldsymbol{\Psi}\boldsymbol{\Psi}^H\mathbf{A}^H = \mathbf{A}\mathbf{A}^H.
		\label{eq:b_simplify}
	\end{equation}
	This simplification decouples the dictionary structure from the orthogonalization process, significantly reducing the optimization complexity.

	\subsubsection{Matrix-Free Implicit Construction}
	To avoid allocating $\mathcal{O}(MN)$ memory for $\mathbf{A}$, we extract $\mathbf{B}$ via black-box impulse probing. 
	Let the forward operator $\mathbf{A}: \mathbb{R}^N \to \mathbb{C}^M$ and its adjoint $\mathbf{A}^*$ be defined compactly as:
	\begin{align}
		\mathbf{A}(\mathbf{c}) &= \mathbf{M} \odot \mathcal{F}(\boldsymbol{\Psi} \mathbf{c}) \\
		\mathbf{A}^*(\mathbf{y}) &= \boldsymbol{\Psi}^H \mathcal{F}^{-1}(\mathbf{M} \odot \mathbf{y})
	\end{align}
	where $\mathcal{F}$ is the 2D-FFT and $\mathbf{M}$ is the undersampling mask. 
	We probe the system with canonical unit vectors $\mathbf{e}_i \in \mathbb{C}^M$:
	\begin{equation}
		\mathbf{B}_{:,i} = \mathbf{A}(\mathbf{A}^*(\mathbf{e}_i)), \quad i = 1, \dots, M
	\end{equation}
	This implicit construction builds the dense $M \times M$ dual matrix without ever instantiating $\mathbf{A}$.

	\subsubsection{Gradient Descent Optimization}
		Inspired by the Gram matrix optimization proposed by Abolghasemi \textit{et al.} \cite{abolghasemi2010} 
		and the system conditioning concepts by Bruckstein \textit{et al.} \cite{bruckstein2008}, 
		we seek a projection matrix $\mathbf{P} \in \mathbb{C}^{M \times M}$ that orthogonalizes the equivalent measurement channels. 
		However, unlike \cite{abolghasemi2010} which optimizes the full $N \times N$ dictionary, 
		our unitary simplification allows us to minimize the Frobenius norm difference directly in the dual measurement space:
		\begin{equation}
			\min_{\mathbf{P}} J(\mathbf{P}) = \|\mathbf{P} \mathbf{B} \mathbf{P}^H - \mathbf{I}_M\|_F^2
		\end{equation}
		Defining the intermediate state $\mathbf{T} = \mathbf{P} \mathbf{B}$, the gradient with respect to $\mathbf{P}$ is analytically derived as:
		\begin{equation}
			\nabla_{\mathbf{P}} J = 2(\mathbf{T} \mathbf{T}^H - \mathbf{I}_M) \mathbf{T}
			\label{eq:grad_p}
		\end{equation}
		To ensure rapid and stable convergence, we dynamically compute the optimal step size $\alpha = 1/L$,
		 where the Lipschitz constant $L$ is estimated via power iteration on the spectral radius of the localized Hessian.
		
	\subsection{Integrated Sparse Reconstruction}

	With both preconditioners established, the effective forward and adjoint operators for the ISTA reconstruction solver are updated to:
	\begin{align}
		\tilde{\mathbf{A}}(\mathbf{c}) &= \mathbf{P} \mathbf{A}(\mathbf{Q} \mathbf{c}) \\
		\tilde{\mathbf{A}}^*(\mathbf{y}) &= \mathbf{Q}^H \mathbf{A}^*(\mathbf{P}^H \mathbf{y})
	\end{align}
	The reconstruction is then iteratively solved via standard soft-thresholding:
	\begin{equation}
		\mathbf{c}^{(k+1)} = \mathcal{S}_{\lambda}(\mathbf{c}^{(k)} - \eta \tilde{\mathbf{A}}^*(\tilde{\mathbf{A}}(\mathbf{c}^{(k)}) - \mathbf{y}))
	\end{equation}
	where $\mathcal{S}_{\lambda}$ is the soft-thresholding operator and $\eta$ is the step size.

	\subsection{Computational Complexity}

	The proposed framework substantially reduces the optimization burden by operating exclusively in the measurement space. 
	The complexity breakdown is summarized in Table~\ref{tab:complexity}.

	\begin{table}[h]
		\centering
		\caption{Computational Complexity Breakdown}
		\label{tab:complexity}
		\begin{tabular}{|l|c|}
			\hline
			\textbf{Operation Phase} & \textbf{Complexity} \\
			\hline
			Implicit $\mathbf{B}$ Extraction & $\mathcal{O}(M \cdot N \log N)$ \\
			$\mathbf{P}$ Gradient Descent ($k$ iters) & $\mathcal{O}( k \cdot M^2)$ \\
			Feature Rotation ($\mathbf{Q}$ matrix) & $\mathcal{O}(N)$ \\
			Integrated ISTA (per iteration) & $\mathcal{O}(N \log N + M^2)$ \\
			\hline
		\end{tabular}
	\end{table}

	For clinical MRI scenarios (e.g., $N = 16,384$, $M \approx 0.2N$), the $\mathcal{O}(M^2)$ cost of optimizing $\mathbf{P}$ 
	is computationally trivial compared to the traditional $\mathcal{O}(N^2)$ dictionary optimization, 
	allowing the algorithm to execute efficiently on standard hardware.

	\section{Theoretical Analysis of Coherence Suppression}
	\label{sec:theory}

	The fundamental bottleneck in highly ill-conditioned sensing matrices, 
	particularly under Cartesian undersampling, 
	is the heavy-tailed distribution of the off-diagonal elements in the Gram matrix 
	$\mathbf{G} = \mathbf{D}^H\mathbf{D}$. 
	This structural alignment implies that a significant fraction of columns in the dictionary 
	$\mathbf{D}$ are nearly indistinguishable,
	 which severely violates the restricted isometry property.

	{
	
	\subsection{Asymptotic Orthogonality of the Random Rotator}
	The fundamental bottleneck in Cartesian-sampled MRI is the heavy-tailed distribution of the Gram matrix. 
	To disrupt this structural alignment, we utilize the DogRot matrix $\mathbf{Q}$ defined in \eqref{eq:q_construct}. 
	The theoretical properties of $\mathbf{Q}$ are summarized as follows (rigorous proofs can be found in \cite{dogrot2026}):

	1) \textit{Spectral Invertibility}: Since $\mathbf{Q} - \mathbf{I}_N$ is a real skew-symmetric matrix, 
	its eigenvalues strictly follow the form $1 \pm cj$ ($c \in \mathbb{R}$), which ensures $\det(\mathbf{Q}) \neq 0$ 
	and thus $\mathbf{Q}$ is invertible almost surely.

	2) \textit{Asymptotic Orthogonality}: In high-dimensional regimes, 
	the mutual coherence between columns of the mixed dictionary follows a sub-exponential distribution. 
	Via Bernstein-type concentration inequalities, the probability of coherence exceeding a threshold 
	$\tau$ decays exponentially, effectively crushing the heavy tail of the coherence distribution.

	3) \textit{Sparsity Preservation}: Due to the diagonal dominance of $\mathbf{Q}$ 
	controlled by $\epsilon$, the signal energy remains concentrated on the original support $\mathcal{S}$. 
	Thus, the transformed signal preserves support identifiability, 
	satisfying $\text{supp}(H_t(\mathbf{Q}^{-1}\mathbf{c})) = \text{supp}(\mathbf{c})$ 
	where $H_t$ denotes the hard-thresholding operator.
	}

	\subsection{Synergistic Coherence Suppression via Measurement Projection}

	While $\mathbf{Q}$ provides a probabilistic structural breakdown of coherence in the feature space, the measurement-space projector $\mathbf{P}$ deterministically suppresses the residual correlations.

	The effective dictionary under the proposed frameworkis $\mathbf{D}_{\text{eff}} = \mathbf{P}\mathbf{A}\boldsymbol{\Psi}\mathbf{Q}$, where $\mathbf{P}$ and $\mathbf{Q}$ are dual-space preconditioners.
	The exact Gram matrix of this augmented system is:
	\begin{equation}
		\mathbf{G}_{\text{eff}} = \mathbf{Q}^H \boldsymbol{\Psi}^H \mathbf{A}^H \mathbf{P}^H \mathbf{P} \mathbf{A} \boldsymbol{\Psi} \mathbf{Q}
		\label{eq:g_eff}
	\end{equation}

	The objective in \eqref{eq:grad_p} ensures that $\mathbf{P}$ acts 
	as a perfect whitening filter in the row space of $\mathbf{A}$.

	Consequently, the inner core of Eq.~(\ref{eq:g_eff}), given by $\boldsymbol{\Psi}^H (\mathbf{P}\mathbf{A})^H (\mathbf{P}\mathbf{A}) \boldsymbol{\Psi}$, 
	behaves as a well-conditioned operator whose non-zero eigenvalues are strictly equalized. 
	The dual-space projection thus acts as a synergistic mechanism:
	\begin{itemize}
		\item \textbf{Deterministic Whitening (Measurement Space):} Actively flattens the spectrum of the dual Gram matrix, 
		ensuring the measurement channels are statistically orthogonal.
		\item \textbf{Probabilistic Rotation (Feature Space):} Randomly rotates the feature-space atoms 
		to prevent any single sparsity basis vector from aligning with the residual null-space of $\mathbf{P}\mathbf{A}$.
	\end{itemize}

	The combination of these two operations rigorously minimizes the worst-case mutual coherence $\mu(\mathbf{D}_{\text{eff}})$, 
	fundamentally expanding the radius of guaranteed exact recovery for the ISTA solver.

	\section{Experimental Validation}
	\subsection{Experimental Setup}
	We evaluate the PAQ framework on clinical MRI datasets covering 
	Glioma, Meningioma, Pituitary, and No-tumor groups, 
	with all images normalized to $128 \times 128$. 
	To simulate severe ill-conditioning, a 20\% Cartesian undersampling mask is applied. 
	Reconstruction is performed using both the classical ISTA and the deep-learning-based VarNet to verify generalizability. 
	We compare two configurations: \textbf{(1) Baseline (A):} Standard ISTA/VarNet reconstruction using the unoptimized Fourier operator and DCT basis; 
	\textbf{(2) Proposed (PAQ):} The synergistic framework incorporating the random rotator $\mathbf{Q}$ ($\epsilon = 0.005$) and the orthogonalization projector $\mathbf{P}$.
	
	\subsection{Quantitative Evaluation}
	Table~\ref{tab:psnr_comparison} and Table~\ref{tab:varnet_psnr_comparison} summarize the PSNR achieved by the ISTA and VarNet solvers.

	\begin{table}[htbp]
		\centering
		\caption{PSNR (dB) Comparison of ISTA Reconstruction under 20\% Cartesian Sampling}
		\label{tab:psnr_comparison}
		\begin{tabular}{lcc}
			\hline
			\textbf{Tumor Type} & \textbf{Baseline ($\mathbf{A}$)} & \textbf{Proposed ($\mathbf{PAQ}$)} \\
			\hline
			Glioma & 14.91 & \textbf{16.15} \\
			Meningioma & 12.51 & \textbf{14.12} \\
			Pituitary & 11.82 & \textbf{13.45} \\
			Notumor & 11.71 & \textbf{13.02} \\
			\hline
		\end{tabular}
		\begin{flushleft}
		\small \textit{\footnotesize Note: Reconstruction time per slice is 0.11s (Baseline) vs 0.16s (PAQ) on Intel i9-13980HX. The pre-computation of $\mathbf{P}$ is an offline process.}
		\end{flushleft}

		{
		
		\centering
		\caption{PSNR (dB) Comparison of VarNet Reconstruction under 20\% Cartesian Sampling}
		\label{tab:varnet_psnr_comparison}
		\begin{tabular}{lcc}
			\hline
			\textbf{Tumor Type} & \textbf{Baseline ($\mathbf{A}$)} & \textbf{Proposed ($\mathbf{PAQ}$)} \\
			\hline
			Glioma & 23.60 & \textbf{23.91} \\
			Meningioma & 20.49 & \textbf{21.64} \\
			Pituitary & 22.02 & \textbf{22.38} \\
			Notumor & 20.02 & \textbf{20.84} \\
			\hline
		\end{tabular}
		\begin{flushleft}
		\small \textit{\footnotesize Note: Inference time per slice is 1.87s (Baseline) vs. 1.99s (PAQ) on NVIDIA RTX 4060. Offline optimization of $\mathbf{P}$ does not affect real-time performance.}
		\end{flushleft}}
	\end{table}
		
	{
	
	As illustrated in Tables~\ref{tab:psnr_comparison} and~\ref{tab:varnet_psnr_comparison}, 
	the unoptimized sensing matrix $\mathbf{A}$ suffers from high structural coherence, 
	which significantly limits the recovery potential of both classical 
	and deep-learning-based solvers. 
	By implicitly reshaping the effective dictionary via the dual-space $\mathbf{PAQ}$ 
	preconditioner, 
	the proposed method consistently yields PSNR improvements regardless of the underlying optimization backbone.

	It should be noted that the proposed framework is a low-complexity, 
	plug-and-play preconditioning approach. 
	Its performance gains are determined by the nature of the sensing problem and the optimization capacity of the underlying solver 
	(e.g., ISTA and VarNet).
	While mainstream deep-learning methods may offer significant overall improvements, 
	the extreme difficulty of $20\%$ Cartesian sampling poses an inherent bottleneck, 
	making further gains limited even with advanced models.
	}

	\subsection{Visual Comparison}
\begin{figure}[htbp]
		\centering
		\subfloat[Ground Truth]{\includegraphics[width=0.33\linewidth]{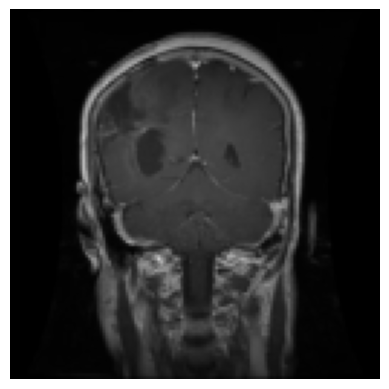}}\hfil
		\subfloat[Baseline ($\mathbf{A}$)]{\includegraphics[width=0.33\linewidth]{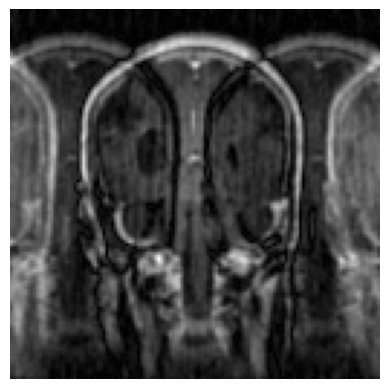}}\hfil
		\subfloat[Proposed ($\mathbf{PAQ}$)]{\includegraphics[width=0.33\linewidth]{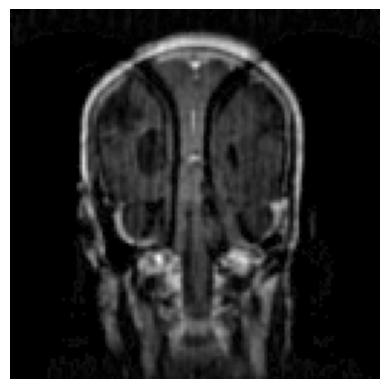}}		
		
		\subfloat[Ground Truth]{\includegraphics[width=0.33\linewidth]{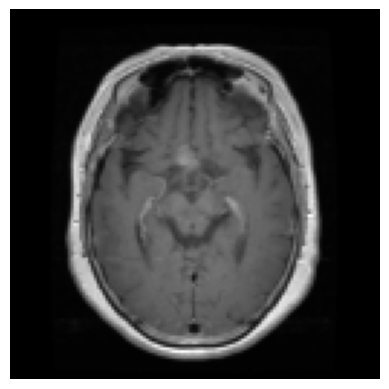}}\hfil
		\subfloat[Baseline ($\mathbf{A}$)]{\includegraphics[width=0.33\linewidth]{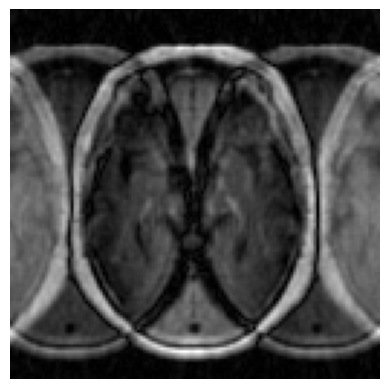}}\hfil
		\subfloat[Proposed ($\mathbf{PAQ}$)]{\includegraphics[width=0.33\linewidth]{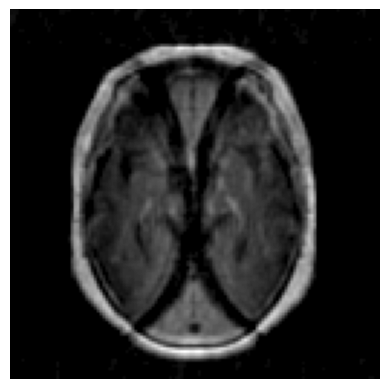}}		
		\caption{\footnotesize Visual comparison under 20\% Cartesian sampling. 
		While baseline (A) suffers from severe structured aliasing, 
		our PAQ effectively restores sharp anatomical features and suppresses artifacts.}
		\label{fig:visual_comparison}
	\end{figure}

	To further demonstrate the effectiveness of the proposed method in suppressing 
	structural artifacts, Figure~\ref{fig:visual_comparison} presents a visual comparison of the reconstructed MRIs.

	Under $20\%$ Cartesian undersampling, the inherent correlation between the Fourier 
	measurements and the DCT basis produces structured aliasing artifacts. 
	The proposed $\mathbf{PAQ}$ framework orthogonalizes the coherent directions, 
	allowing the ISTA solver to isolate the sparse coefficients more accurately, 
	which is evidenced by the visibly cleaner reconstructions and reduced background noise.

	\section{Conclusion}
	
	This paper proposes a novel approach to coherence minimization in compressed sensing MRI through joint optimization of 
	measurement-space ($\mathbf{P}$) and feature-space ($\mathbf{Q}$) projection matrices. 
	The key contribution is the \textbf{dual-space} formulation of $\mathbf{P}$ optimization, 
	which reduces computational complexity from $\mathcal{O}(N^2)$ to $\mathcal{O}(M^2)$ while maintaining theoretical 
	soundness. By combining implicit matrix-free operators (DCT/FFT), black-box impulse probing for 
	dual matrix construction, and adaptive gradient descent with convergence safeguards, we achieve 
	a practical and scalable algorithm suitable for clinical deployment.
	
	Experimental validation on glioma MRI data demonstrates that the proposed $\mathbf{PAQ}$ optimization method 
	yields measurable reconstruction quality improvements compared to baseline and Q-only methods, 
	while maintaining computational efficiency. 
	
	\begingroup
	\scriptsize
	\bibliographystyle{IEEEtran} 
	\bibliography{references.bib}
	\endgroup

\end{document}